\documentclass[bibyear]{aa} 

\usepackage{graphicx}
\usepackage{txfonts}
\begin{document}

   \title{The origin of the Far-infrared continuum of $z\sim 6$ quasars}

   \subtitle{A radiative transfer model for SDSS J1148+5251}

   \author{R. Schneider\inst{1}\thanks{raffaella.schneider@oa-roma.inaf.it},
           S. Bianchi\inst{2},
           R. Valiante\inst{1},
           G. Risaliti\inst{2}
           \and
           S. Salvadori\inst{3}           
          }

   \institute{$^1$INAF/Osservatorio Astronomico di Roma, Via di Frascati 33, 00090 Roma, Italy\\
              $^2$INAF/Osservatorio Astrofisico di Arcetri, Largo Enrico Fermi 5, 50125 Firenze, Italy\\
              $^3$Kapteyn Astronomical Institute, Landleven 12, 9747 AD Groningen, The Netherlands \\
             }

   \date{ }

  \abstract
   {Understanding the history of formation of $z > 6$ quasars represents a major challenge to theoretical
   models. Physical insights on the connection between the central black hole and its host galaxy can be
   gained by means of the quasar infrared properties.
   }   
   {Here we investigate the origin of the far-infrared continuum of SDSS J1148+5251,
    using it as a prototype for the more general class of high-luminosity high-redshift quasars.
   }
   {We run the radiative transfer code TRADING to follow the transfer of radiation from the central source and 
   from stellar sources through the dusty environment of the host galaxy. We adopt
   simple models for the central source,
   including all the radiation that can travel beyond the dusty torus. The radiation from stellar sources
   is modeled using the code P{\'E}GASE.
   The model is based on the output of the semi-analytical 
   merger tree code, \textsc{GAMETE/QSOdust}, which enables to predict the evolution of the
   host galaxy and of its nuclear black hole, following the star formation history and
   chemical evolution -- including dust -- in all the progenitor galaxies of SDSS J1148+5251. 
   }
   {We find that the radiation emitted by the central source, which dominates the observed spectral
   energy distribution from UV/optical to near and mid infrared wavelengths, can also provide an
   important source of heating for the dust distributed in the host galaxy, powering at least 30\% and up to
   70\% of the observed far infrared emission at rest-frame wavelengths $[20 - 1000]\, \mu$m. 
   The remaining fraction is contributed by stellar sources and can 
   only be achieved if the host galaxy is able to sustain a star formation rate 
   of $\rm \approx 900 \, M_{\odot}/yr$ at $z=6.4$. This points to a co-evolution scenario where, during their
   hierarchical assembly, the first
   super-massive black holes and their host galaxies first grow at the same pace until the black hole
   reaches a mass of $\rm \sim 2 \times 10^8 M_\odot$ and starts growing faster than its host, 
   reaching the bright quasar phase when the black hole and stellar mass fall within
   the scatter of the scaling relation observed in local galaxies.  
   This same evolutionary scenario has been recently shown to explain the properties of a larger sample of $5 < z <6.4$ QSOs,
   and imply that current dynamical mass measurements may have missed an important fraction of the host
   galaxy stellar mass.
   }
   {We conclude that the far-infrared luminosity of high-z quasars is a sensitive tracer of the rapidly changing physical conditions
   in the host galaxy. Quasars appear far-infrared bright when the host galaxy
   can still sustain strong starbursts, with star formation rates $\rm 100 \, M_\odot/yr < SFR < 1000 \, M_\odot/yr$ and progressively 
   dim as large quasar-driven outflows deplete the host galaxy of its gas content, damping star formation and leaving
   the central source as the only source of dust heating.}


   \maketitle

\section{Introduction}
\label{sec:intro}

%

The first comprehensive far-infrared (FIR) investigations 
on optically selected low-$z$ AGNs have been made using 
{\it Infrared Astronomical Satellite} ({\it IRAS}) and 
{\it Infrared Space Observatory} ({\it ISO}) data, supplemented with SCUBA and IRAM data at longer wavelengths
(see e.g. Haas et al. 2000, 2003). A variety of different Spectral Energy Distribution (SED) shapes was observed, suggesting that
the mid-infrared (MIR) and FIR emission originate from dust components whose heating mechanisms
may be different (Haas et al. 2000). In particular, while in all sources the MIR emission could be powered by the central
Active Galactic Nuclei (AGN), the relative role of starbursts and AGN in powering the FIR emission was hard to
established, although Haas et al. (2003) proposed that QSOs with the largest FIR luminosities ($\rm L_{FIR} \ge 10^{13} - 10^{14} L_\odot$)  
favour a strongly dominating AGN contribution. 

Progress in the field was enabled by observations with the {\it Spitzer Space Telescope}, using 
MIR polycyclic aromatic hydrocarbon
(PAH) emission features as tracers of star formation in QSO host galaxies (Schweitzer et al. 2006; 
Netzer et al. 2007;  Lutz et al. 2008).  
Indeed, the main MIR PAH emission features at 6.2, 7.7, 8.6, and 11.3 $\rm \mu m$  have been 
extensively used as extragalactic star formation indicators because of their presence
in star forming environments with a wide range of physical conditions and their absence in the
proximity of AGNs. While the calibration between the PAH luminosity and the star formation
rate depends on the physical properties of the interstellar medium, PAH emission is sufficiently
strong to be less easily outshone by the AGN emission, unlike $\rm H\alpha$ or the 24 $\mu$m  continuum.
Netzer et al. (2007) applied this technique to local QSOs; comparing the average SEDs of groups of QSOs  
 that are differing in their FIR luminosity, they were able to derive a SED for the pure AGN, after subtraction
of the host star formation.
Lutz et al. (2008) have extended the study to a sample 
of mm-detected QSOs at $1.8 \le z \le 2.8$, showing that these sources follow the same correlation between the PAH luminosity 
and the rest-frame FIR emission inferred for local QSOs. For mm-bright QSOs, this provides strong evidence for intense star
formation rates in their hosts, sometimes exceeding 1000 $\rm M_\odot/yr$. However, the mm-faint end of this QSO population 
may contain pure AGNs, with no strong host star formation, and the observed weak rest frame FIR emission appears consistent with
pure AGN-heated dust (Lutz et al. 2008).

The launch  of the {\it Herschel Space Observatory}  has enabled to directly observe the galaxy rest-frame FIR emission. 
Using large samples of type 1 and type 2 AGN in the {\it Herschel} Multi-tiered Extragalactic Survey (HerMES)
fields observed with SPIRE at  $\rm 250, 350, \, and \,500 \, \mu$m, Hatziminaoglou et al. (2010) 
showed that quasars out to $z \sim 4$ have the same FIR colours as star-forming galaxies 
and that a starburst was always needed to reproduce the FIR emission (see also Elbaz et al. 2010).
By matching a sample of 24\,$\mu$m - selected QSOs in the {\it Spitzer} Wide-area InfraRed Extragalactic Survey
to HerMES, Dai et al. (2012) collected 32 QSOs with
$0.5 \le z < 3.6$ with SPIRE data. These FIR-detected QSOs show
cold dust emission with $\rm L_{FIR}$ in the range $\rm10^{11.3} - 10^{11.5} \, L_\odot$, dust masses
of $\rm 10^8 - 10^9 \, M_\odot$ and temperatures from 18 K to 80 K, similar to ULIRGs and (sub)mm
detected QSOs. They find that the FIR properties can not be predicted from shorter
wavelengths,  because at rest-frame frequencies $0.3 - 20\, \mu$m FIR-detected
QSOs have similar SEDs of FIR-undetected QSOs at comparable redshifts. Assuming that
star formation is powering the FIR emission, about 40\% of the sample requires $\rm SFR > 1000 \, M_\odot/yr$; yet, 
for some QSOs the derived SFRs 
are $\rm > 5000\, M_\odot/yr$, which is unlikely and may imply a non-negligible contribution from the AGN. 
Multi-components SED fits of a much larger sample of 250 $\mu$m selected galaxies of HerMES with {\it Spitzer}/InfraRed Spectrograph
spectra show, however, that the presence of a starburst is necessary to account for the total FIR emission, and that 
the presence of an AGN has no noticable effect on the FIR properties of the host galaxy (Feltre et al. 2013).

Similar studies have been recently applied to 69 QSOs at $z > 5$ by Leipski et al. (2014),
who have presented SEDs covering the
rest-frame wavelengths from $0.1\, \mu$m to $\sim 80 \, \mu$m, including {\it Spitzer} and {\it Herschel} 
observations. 
At rest-frame wavelengths $\rm \ge 50 \, \mu m$, the SEDs of the 10  sources which have been detected in at 
least 4 of the 5 {\it Herschel} bands (with PACS at 100 and 160 $\mu$m, and with SPIRE at $\rm 250, 350, and \, 500\,\mu$m) are completely dominated by a luminous cold
dust component with $\rm 47 \, K  \le  T_{FIR} \le 60 \, K$ and $\rm L_{FIR} \sim 10^{13} \, L_\odot$. 
The origin of the FIR component is further analyzed by
stacking the SEDs of {\it Herschel}-detected (10 objects detected at $\rm 160, 250, and \, 350 \,\mu m$), 
partly {\it Herschel}-detected (14 objects detected at $\rm 100\,  and/or \, 160 \, \mu m$) and 33 {\it Herschel}-non
detected sources (33 objects). At rest-frame wavelengths $\rm \lambda \le 15 \, \mu m$, the average SEDs of the 
{\it Herschel}-detected and partly {\it Herschel}-detected sub-samples appear very similar,  but partly detected sources show
a substantial drop in their average SED at longer wavelengths. 
The fact that some optical and MIR luminous QSOs do show FIR
emission and others do not is interpreted as an indication that star formation is the dominant driver for the additional
FIR component observed in {\it Herschel}-detected QSOs. 


All the above studies adopt physically motivated multi-component SED fitting techniques
which - however - miss the energy balance between the various components. A notable exception is
the work by Dwek, Galliano \& Jones (2007), who have made a simple decomposition of the observed SED
of SDSS J1148 into different components. They used a simple screen
model with a Galactic extinction law to calculate
the spectrum of the escaping stellar radiation, choosing the magnitude of the extinction
so that the total reradiated FIR emission was equal to the total energy absorbed by the dust.
In this way they estimated that the inferred FIR luminosity of $\rm 4.6 \times 10^{13} L_\odot$ could be 
powered by star formation but requires a continuous star formation rate of 2500 $\rm M_\odot/yr$ for 400 Myr
with a top-heavy Initial Mass Function (IMF).

Here we attempt to make a step forward 
by using a detailed radiative transfer (RT) model to follow the contribution of the host star 
formation and the central AGN to dust heating.  Our aim is to 
 to assess to what extent the observed SEDs of 
$z \sim 6 $ QSOs in the rest-frame FIR can place constraints on current scenarios
for the formation and evolution of the first Super-Massive Black Holes (SMBHs) and their host galaxies. 

Information on the star formation rate, the 
nature of the stellar populations, the mass of gas and dust of the host galaxy are obtained 
using the data-constrained semi-analytical merger tree model, \textsc{GAMETE/QSOdust}
(Valiante et al. 2011, 2012, 2014). 
This model allows to follow the star formation history
and the mass growth of the nuclear Black Hole (BH) in all the progenitor galaxies that contribute to
the hierarchical build-up of $z \sim 6 $ QSOs. Valiante et al. (2011) show that when
the model is applied to SDSS J1148+5251 (hereafter SDSS J1148), one of the best-studied 
QSOs at $z = 6.4$,  the observed properties of the QSO (BH, gas and dust masses, SFR,
stellar mass) are reproduced provided that distinct evolutionary paths are followed. This
same conclusion has been recently confirmed to hold for a larger sample of
QSOs at $5 < z < 6.4$ (Valiante et al. 2014) which share a common formation
scenario: in all the progenitor galaxies of the final host dark matter halo,
stars form according to a standard, Salpeter-like IMF via quiescent
star formation and efficient merger-driven bursts. At the same time, the central BH
grows via gas accretion and mergers with other BHs. When the BH mass reaches a threshold
value of $\rm 2 \times10^8 - 10^9 \, M_\odot$, a strong energy-driven wind starts to clear-up 
the interstellar medium of dust and gas through a large outflow, damping the star formation rate 
and un-obscuring the line of sight toward the QSO. Although the estimated outflow rates
can be as large as a few $\rm 1000 \, M_\odot/yr$, in agreement with current observational 
constraints (Maiolino et a. 2012; Cicone et al. 2015), the final star formation rates are
comparably large, $\sim 600 - 2000 \, \rm M_\odot/yr$, due to intense merger-driven bursts. 
As a result, all the QSOs host galaxies are characterized by final stellar masses in the range 
$\rm (4 - 6)\times 10^{11} M_\odot$, a factor 3 - 30 larger than the upper limits allowed 
by the observed dynamical and molecular gas masses. 

Although large uncertainties still affect dynamical mass measurements in these high-$z$ galaxies,
one possible solution to this so-called {\it stellar mass crisis} would be to assume that the cold
gas is converted into stars at a smaller efficiency than adopted by the scenario described above. In this
way, the BH grows faster than the host galaxy and when it reaches a final mass of $\rm \sim 10^9 \, M_\odot$,
the stellar mass is still $\rm 10^{10} - 10^{11} M_{\odot}$ (Valiante et al. 2014). Yet, in order to reproduce 
the observed chemical properties of the host galaxy, in particular the dust mass, the stars must form 
according to a top-heavy IMF, with a characterstic mass of $\rm 5 M_\odot$, that allows to maximize 
the stellar metal and dust yield (Valiante et al. 2011). More importantly, the final SFRs are 
$\rm < 100 \, M_{\odot}/yr$, significantly smaller than in the standard model. 

It is clear that the two scenarios discussed above result in very different predictions on the
relative role of the AGN and star formation as the powering sources of the FIR emission. Hence,
any attempt to gain a deeper understanding of the origin of the FIR emission in high-$z$
QSOs may provide important insights on the co-evolution of the first SMBHs and their host galaxies. 

The paper is organized as follows: in section \ref{sec:model} we give a short description of
the main features of the semi-analytical model  \textsc{GAMETE/QSOdust} and we 
present the standard scenario for SDSS J1148 at $z= 6.4$, using this QSO as a prototype
for the general class of high-$z$ QSOs; 
in section~\ref{sec:sed} we describe the numerical set-up and the radiative
transfer model; in sections ~\ref{sec:central} - \ref{sec:dust}, we present the 
adopted model for the spectrum emitted by the central AGN and by the stars,
and the assumed spatial distribution of the dust in the host galaxy; in section
\ref{sec:results} we illustrate and critically discuss our main results and in 
section \ref{sec:conclusions} we draw our conclusions.     

For consistency with previous works (Valiante et al. 2011, 2014), 
we adopt a Lambda Cold Dark Matter ($\Lambda$CDM) 
cosmology  with $\Omega_m = 0.24$, $\Omega_\Lambda = 0.76$, $\Omega_b = 0.04$, 
and $H_0 = 73$~km/s/Mpc. 
The age of the Universe at a redshift $z = 6.4$ is 900 Myr and the 
model and observations are scaled using the luminosity distance at this redshift, 64.5 Gpc.

\section{Modeling the evolution of high-redshift quasars}
\label{sec:model}

In this section we give a brief presentation of the best-fit models for the
formation and evolution of the QSO J1148 obtained by means of 
the semi-analytical code \textsc{GAMETE/QSOdust} (Valiante et al. 2011, 2012, 2014). 
We focus here only on those features which are particularly relevant for
the present study and address the interested reader to Valiante et al. (2011)
for a comprehensive description of the code.\\  
\\
\noindent
{\it Dark matter and BH evolution} \\
High-$z$ QSOs are likely to reside in dark matter (DM) halos with masses  $\sim \rm 10^{12} - 10^{13} \, M_\odot$
(Fan et al. 2004; Volonteri \& Rees 2006; Valiante et al. 2012; Fanidakis et al. 2013).  Here we assume that J1148
is hosted by a $\rm 10^{13}~M_\odot$ DM halo at $z = 6.4$ whose evolution, backward in time, is simulated using
a binary Monte Carlo merger tree algorithm (Valiante et al. 2011).
To save computational time, we only resolve dark matter 
halos with a virial temperature $\rm T_{vir} > 10^4~K$ that can cool through
hydrogen Lyman-$\alpha$ emission; in the first star forming halos 
we plant BH seeds with a mass of $10^4 h^{-1} \rm M_{\odot}$ and stop
populating newly virialized halos once their mass falls below the mass of a 
$4\sigma$ halo at the same redshift. We then follow the subsequent evolution
of these BH seeds as they grow through mergers and gas accretion. 
In our formulation, gas accretion onto the central BH is Eddington-limited
and regulated by the Bondi-Hoyle accretion rate, with a constant 
efficiency parameter, $\alpha \sim 200$, set to reproduce the observed BH
mass of J1148 of  $\rm \sim 3 \times 10^{9}~M_{\odot}$ 
(Barth et al. 2003; Willott, McLure and Jarvis 2003).\\

\noindent
{\it BH feedback and outflow rates}\\
The gas content within each progenitor galaxy is regulated by infall and outflow
rates as well as by the rate of star formation and of returned gas at the end of
stellar evolution. Gas outflows can be powered by supernova explosions and BH
feedback; the relative importance of these two physical processes depends on
the evolutionary phase of each progenitor galaxy. At redshifts $6.4 < z < 8$, the
dominant effect is  AGN feedback, since BH-driven outflows have already
significantly depleted the gas content of the host galaxy, leading to a down-turn
of the star formation rate (see Figs. 3 and 4 in Valiante et al. 2012). 
It is important to stress that, despite the very schematic implementation of AGN feedback -
described as energy-driven winds - the predicted outflow rate at $z=6.4$ is in good 
agreement with the massive gas outflow rate of $\rm \sim 3500~M_{\odot}/yr$ 
detected by Maiolino et al. (2012) and Cicone et al. (2015) through IRAM
PdBI observations of the [CII] 158~$\mu$m line in the host galaxy of J1148 
(Valiante et al. 2012). \\
\\
\noindent
{\it Star formation law} \\
The evolution of the stellar content of the host galaxy has been investigated 
adopting different prescriptions for the SFR. The star formation rate can occurr
through a quiescent mode, by converting a given fraction of cold gas into stars
on the galaxy dynamical time, or by means of merger-induced starbursts with 
enhanced efficiencies, in the so-called bursting mode (see Valiante et al. 2011).
The relative importance of these two modes of star formation depends on the
mass ratio of the progenitor merging galaxies, $\mu$. When $\mu \rightarrow 1$,
in equal-mass mergers, the starbursts are maximally efficient and the star
formation rate can be up to 80 times greater than in the quiescent mode (see
Table 2 in Valiante et al. 2011 for the numerical values of the parameters).
All the models that we have explored predict a tight co-evolution of the
nuclear BH and stellar content of the host galaxies in the first 600-700 Myr.
However at $z < 8$, when the mass of the nuclear BH has grown to $\rm \sim 10^8~M_{\odot}$ 
and AGN feedback starts to affect the gas content of the host galaxy, the
rate of star formation is very sensitive to the adopted star formation efficiency.
As a result, the final star formation rates and stellar masses can increase from 
$\rm SFR \sim 70 \, M_\odot/yr$ and  $\rm M_{\star} \sim 3 \times 10^{10}~M_{\odot}$
-- for the models which adopt the lowest star formation efficiency -- to $\rm SFR \sim 900 \, M_\odot/yr$ and
$\rm M_{\star} \sim 4 \times 10^{11} \,M_{\odot}$, when the largest star formation efficiency is considered. 
In the former case, the final stellar mass is within the limits allowed by measurements
of the dynamical mass within the central $\rm 2.5~kpc$ of the host galaxy obtained by means 
of the resolved CO emission (Walter et al. 2004). Conversely, in the latter case 
the final stellar mass is closer to the value that would be implied by the 
nuclear SMBH mass of J1148 if the local BH-stellar mass relation were assumed to hold at these very high
redshifts (see Figure 3 in Valiante et al. 2011). We will return to this point in Section \ref{sec:conclusions}.\\

\begin{figure*}
\includegraphics[width=\hsize]{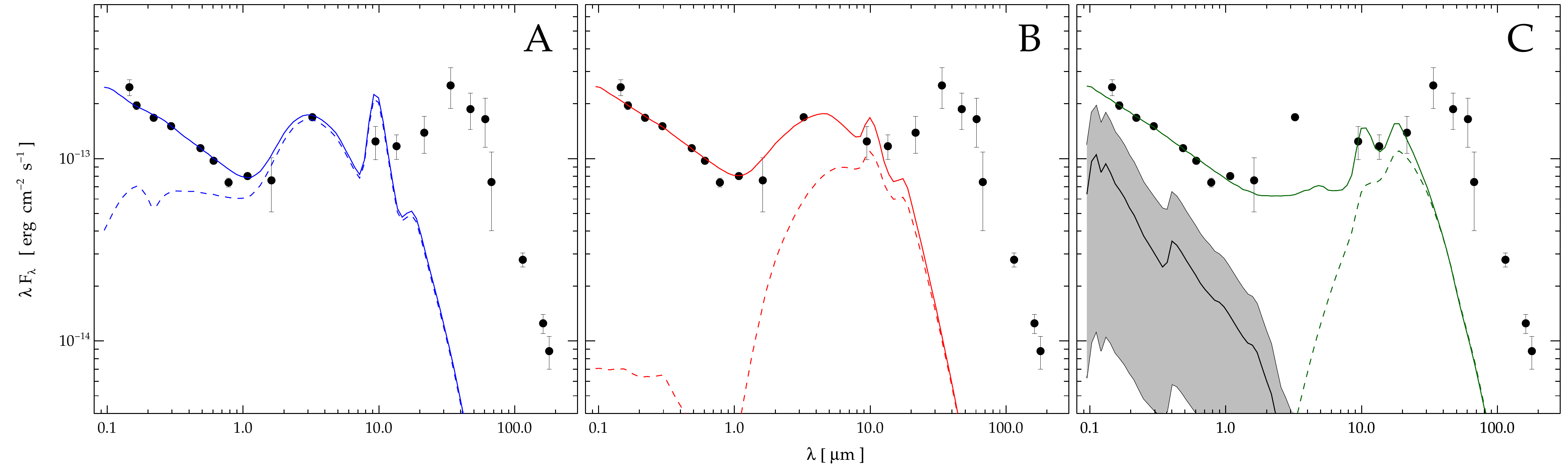}
\caption{The rest-frame SED of J1148 compared to the SED of the central 
source adopted in our models (solid lines for the face-on view, dashed lines 
for the edge-on view). Model A SED is from model {\tt t0.1\_p1\_q6\_oa50\_1pC11\_8}
of Stalevski et al.\ (2012); models B and C are from models 
{\tt N2.5\_a-2.00\_theta60\_Rout150\_tauV80} and {N2.5\_a-0.50\_theta60\_Rout150\_tauV80},
respectively, of H\"onig \& Kishimoto (2010). For clarity, we show the SED of the stars
only in the right panel (shaded region). This is computed with P{\'E}GASE using the output 
of \textsc{GAMETE/QSOdust} standard IMF model (see section 5).}
\label{fig_temp}
\end{figure*}

\noindent
{\it Chemical enrichment}\\
The enrichment of the ISM is followed taking into account the release of heavy elements
and dust grains by Asymptotic Giant Branch (AGB) stars and Supernovae (SN), assuming mass and metal-dependent yields (Valiante et
al. 2009). We do not adopt the instantaneous recycling approximation, hence we take
into account stellar evolution on the characteristic stellar lifetimes.
The mass of dust is computed following the cycling of dust in the interstellar medium,
taking into account grain destruction in the hot diffuse phase of the interstellar medium
by interstellar shocks and grain-growth in dense
molecular clouds (de Bennassuti et al. 2014; Valiante et al. 2014). 
Indeed, Valiante et al. (2011) show that for the adopted dust yields, stellar sources can produce up to 
$\approx \rm 3 \times 10^7~M_{\odot}$ of dust, as a result of the balance between dust production and 
dust destruction by the combined effect of astration and interstellar SN shocks. This is approximately
a factor ten smaller than the dust mass inferred by fitting the observed rest-frame 
FIR-emission.\footnote{Assuming that the dust is optically 
thin and that it radiates as a 'grey-body', the mass of dust can be computed as, 
\[
M_{\rm dust} = \frac{S_{\nu_0} \, d_{\rm L}^2(z)}{(1+z) \, \kappa_{\rm d} (\nu) \, B(\nu, T_{\rm d})},
\]
\noindent
where $\kappa_{\rm d}(\nu) = \kappa_0 (\nu/\nu_0)^\beta$ is the opacity coefficient per 
unit dust mass, $B(\nu, T_{\rm d})$ is the Planck function for a dust temperature 
$T_{\rm d}$, and $d_{\rm L}$ is the luminosity distance to the source.
Using a $\chi^2$ fit to the observed points at $\lambda \ge 160 \, \mu m$ (see section \ref{sec:sed}) and the average dust properties
of Draine (2003),  which we adopt for the present study, we estimate a dust mass $\rm M_{\rm dust} = 3.6 \times 10^8 M_\odot$,
a dust temperature $\rm T_{d} = 56 K$, and a total FIR luminosity $\rm L_{FIR,obs} = 3.3 \times 10^{13} L_\odot$.
}

Efficient grain growth in dense interstellar clouds must be invoked to reproduce the observed dust mass
(Micha{\l}owski et al. 2010; Valiante et al. 2011).
This mechanism requires, however, that enough gas-phase metals are available to fuel grain growth (de Bennassuti et al. 2014;
Valiante et al. 2014). \\

\noindent
{\it Selected models}\\
The detailed analysis done by Valiante et al. (2011) convincingly showed that if the stars
are formed with a Larson IMF,
\[
\rm \phi(m) \propto m^{-2.35} e^{-m_{ch}/m},
\]
\noindent
the chemical properties
inferred from observations of J1148 host galaxy require that either ({\it i})  the stars are formed with a characteristic stellar mass of $\rm m_{ch} = 0.35~M_{\odot}$ (standard IMF) and with high-efficiency  
or that ({\it ii}) the stars are formed with a characteristic stellar mass of $\rm m_{ch} = 5~M_{\odot}$ (top-heavy IMF) 
but with low-efficiency, 
(see Table 2 and Figure 7 in Valiante et al. 2011). 

As we have discussed above, the two models predict different star formation rates and UV luminosities,
hence FIR emission. 
Using the population synthesis code P{\'E}GASE (see section \ref{sec:stars} and references therein)
and assuming that the stars form in a 50 Myr burst with solar metallicity,
we estimate that ({\it i}) a SFR of $\rm 1 \, M_\odot/yr$ leads to a FIR luminosity in the wavelength range
$\rm [8 -1000] \, \mu$m of  $\rm 1.08 \times 10^{10} L_{\odot}$ 
for $\rm m_{ch} = 0.35 \, M_\odot$ and ({\it ii}) of $\rm 2.89 \times 10^{10} L_{\odot}$ for $\rm m_{ch} = 5 \, M_\odot$.
Although the star formation history that characterize the evolution of the simulated QSO is more
complex and can not be approximated by a single 50 Myr burst, these conversion factors can 
provide a rough estimate of the FIR luminosities contributed by the stars. Given the predicted 
SFRs at $z = 6.4$, we find $\rm L_{FIR,\star} \sim 10^{13} \, L_\odot$ 
and $\rm L_{FIR,\star} \sim 2 \times 10^{12} \, L_\odot$ for the standard and top-heavy IMF models,
respectively. In the same wavelength range, a $\chi^2$ fit to the observed points of J1148 (see Fig.\ref{fig_temp}) 
leads to a luminosity
of $\rm L_{FIR,obs} = 3.3 \times 10^{13} \, L_\odot$  (see footnote 1), close to that
obtained by star formation in the host for a standard IMF model. 
Hence, in what follows we will adopt the standard IMF model  as input to the radiative transfer 
calculations and discuss in section~\ref{sec:results} the implications of our results for the other model.

\section{Modeling the Spectral Energy Distribution}
\label{sec:sed}

The observed SED of J1148 is shown in Fig.~\ref{fig_temp}. Data points come from 
the Sloan Digital Sky Survey (Fan et al. 2003), Subaru telescope (Iwamuro et al. 2004), {\it Spitzer} (Jiang et al. 2006; Hines et al. 2006) as 
reported in Table 1 of Leipski et al. (2010), to which flux densities at 70, 100, 
160, 250, 350 and 500~$\mu$m from {\it Herschel}  were added, fully 
sampling the peak of the rest-frame FIR emission (Leipski et al. 2013).

In the optical, the SED shows the typical power-law (close to $F_\nu \propto \nu^{0.5}$)
of unreddened type-1 QSO, emitted by the accretion disk around the central BH.
Thus the QSO can be considered as seen perpendicularly to the 
dusty torus surrounding the accretion disk (i.e. it is seen {\em face-on}). This
dust-free view can be used to estimate the bolometric luminosity of the 
accretion disk, by
using a template SED that extends outside of the optical wavelengths: we found
a bolometric luminosity 
of $1.6\times 10^{14}$ L$_{\sun}$ for the intrinsic SED
used in the radiative transfer models of Stalevski et al.\ (2012).

The radiation travelling along directions close to {\em edge-on} passes through the 
torus and heats the dust to high temperatures, producing thermal radiation in the
NIR. The mean QSO SED derived by Richards et al.\ (2006) for type-1 QSOs observed 
by SDSS and {\it Spitzer} indeed shows a NIR bump, but this template is unable to predict 
the larger output observed in this wavelength range from high-luminosity, high-redshift 
QSOs such as J1148 (see Fig.\ 7 in Leipski et al. 2013). Beyond 20~$\mu$m, the emission comes
from relatively colder dust in the QSO's host (outside of the dusty torus). As discussed
in Section \ref{sec:intro}, the FIR emission is generally 
believed to be heated by star formation in the host; however, we can not exclude
that the radiation emitted by or escaping from the dusty torus of the QSO
can provide an additional, non negligible contribution. Indeed, this is what we 
intend to assess with a detailed radiative transfer calculation. 

We model the SED using the 3-D Monte Carlo Radiative Transfer code 
TRADING (Bianchi 2008). The code has been developed to study dust extinction 
and emission in dusty spiral galaxies (see, e.g. Bianchi \& Xilouris 2011; Holwerda et al. 2012),
but it has a general structure and can be easily adapted to other astrophysical contexts.
In particular, the code uses an adaptive grid to sample the dust distribution at
different scales in environments of different densities. Despite this ability, a
full radiative transfer model from the sub-pc scale of the dusty torus up to the kpc
scales of the host galaxy is still computationally too expensive. Thus, we are
forced to make a few simplyifing assumptions:\\
\noindent
 {\it (i)} For the central source, we adopt the SED obtained by dedicated, external, 
models for the transfer of the radiation from the accretion disk through the dusty 
torus. We will present the three different models 
 that we have explored in Section \ref{sec:central}. \\
\noindent
{\it (ii)} For the stellar sources in the host galaxy, we adopt the SED computed 
with the P{\'E}GASE population synthesis model using as input the star formation histories, 
age and metallicities of the stellar populations predicted by \textsc{GAMETE/QSOdust}.
Similarly,  \textsc{GAMETE/QSOdust} allows to estimate the mass of gas and dust
in the host. The stellar SED and the adopted spatial distributions of star, gas and dust
are presented in Sections \ref{sec:stars} and \ref{sec:dust}. \\
\noindent
{\it (iii)} Using TRADING, we follow the transfer of radiation from the central source
(accretion disk and dusty torus) and from stellar sources
through the dusty environment of the host galaxy. As customary for RT models in a dusty galaxy, we consider
the transfer of radiation only for $\lambda > 912$\AA\ assuming that ionizing radiation
will be intercepted entirely by the neutral hydrogen in the host.

\begin{figure*}
\center
\includegraphics[width=9cm]{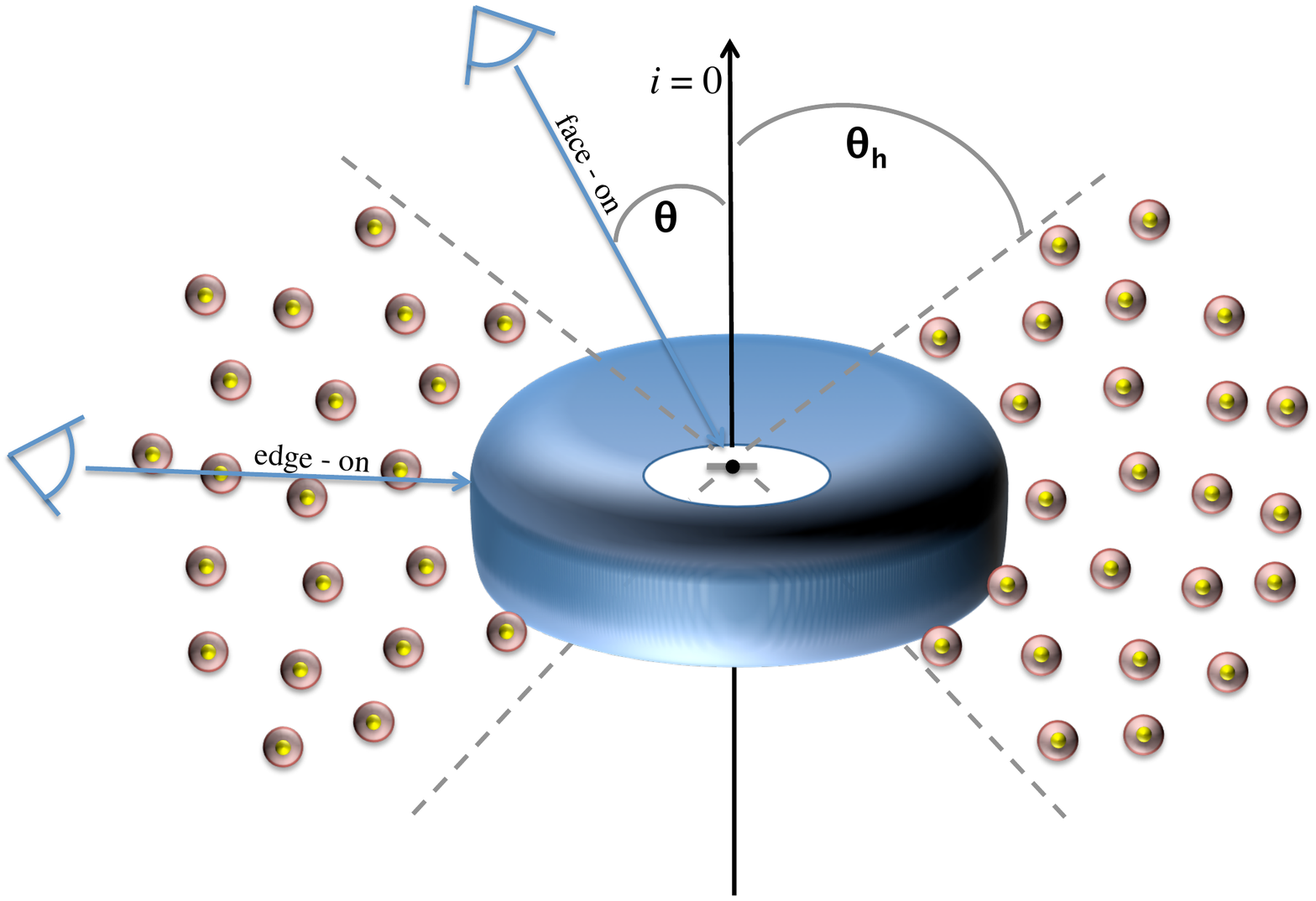}
\includegraphics[width=9cm]{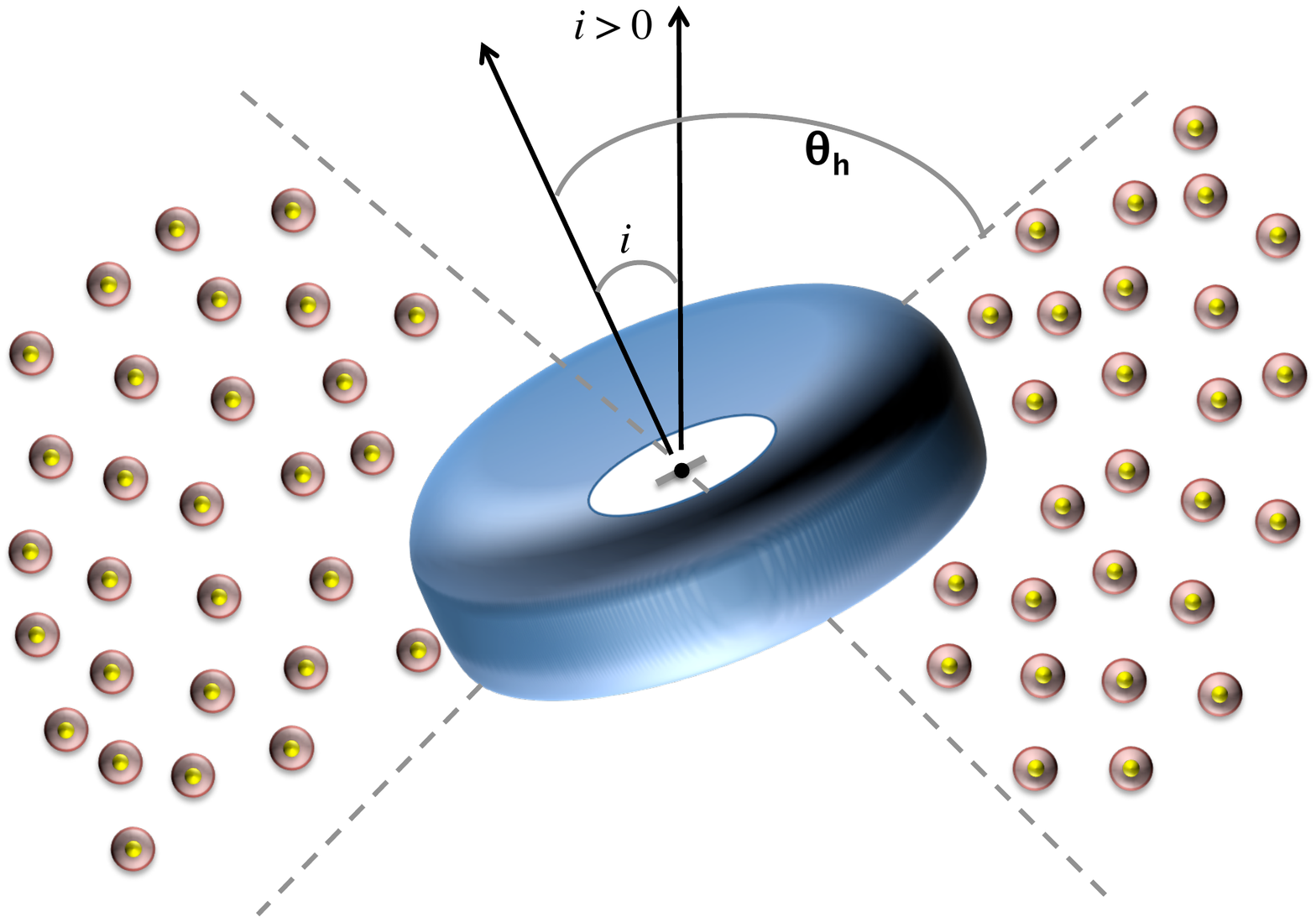}
\caption{Schematic representation of the adopted geometry, assuming two different inclination angles between
the polar axis of the torus and of the host galaxy (${\it i}=0$, left panel, and ${\it i}=20^\circ$, right panel).}
\label{fig:scheme}
\end{figure*}

\section{The central source}
\label{sec:central}

Several models for the radiation emitted by the dusty torus that surrounds the 
accretion disk are available in the literature. These often rely on different 
methods and might result in contradictory and/or non-unique solutions when fitted 
to the data (see H\"onig 2012 for an overview).
In this work, 
our intent is not to determine  the internal properties of the torus but
rather to explore central source models which provide
a reasonable agreement with the observed SED at NIR/MIR wavelengths and study their impact on the FIR emission. 
To this aim we have selected three alternative models extracted from two different libraries available in the
literature: a model from Stalevski et al. (2012) that was computed using a Monte Carlo radiative transfer code similar
to TRADING (model A), and two models from the library of Hoenig \& Kishimoto (2010). The first (model B) provides an
equally good agreement to the observed NIR emission as model A, but with different  physical and geometrical torus
properties. The second (model C) is similar to the SED that has been recently used to fit the MIR emission of 
a sample of high-z FIR-luminous QSOs, including J1148, by Leipski et al. (2013). In addition, these two classes of
models predict very different dusty torus optical depths and enable us to explore the implications of this property to the
contribution of the central source to dust heating in the host galaxy. \\
 
 \noindent 
{\it Model A:}\\
\noindent
The first SED is taken from the simulations of Stalevski et al.\ (2012) that
were computed using the radiative transfer code SKIRT (Baes et al. 2011)\footnote{Stalevski et al.\ (2012) models are available 
at: {\tt https://sites.google.com/site/skirtorus}}. The torus is simulated as
a two-phase medium with half-opening angle of $\theta_h = 40^\circ$, so that 
dust is present for $40^\circ\le \theta \le 140^\circ$, with $\theta$ the polar 
angle from the torus vertical axis. When the torus is seen at inclinations 
$\theta < \theta_h$, the light from the accretion disk is seen unattenuated (see 
the schematic representation in Fig.~\ref{fig:scheme}).  Since these models 
include the radiation directly emitted from the accretion disk, it is 
straightforward to normalize the dust-free {\em face-on} SEDs to 
the observed optical data of J1148. The database was then searched for template 
spectra that best matched the infrared data. It was found that models that 
match the observed data at $\rm \lambda > 10 \, \mu\mathrm{m}$
result in  too high dust emission at shorter wavelengths. Conversely, a few dusty
torus models with relatively low optical depth can roughly reproduce 
the observations at $1\, \mu\mathrm{m} \le \lambda \le 10 \, \mu\mathrm{m}$. 
The model selected 
is the one which passes through the data at 3.2\,$\mu$m without exceeding
the observed flux at wavelengths $\lambda > 10 \, \mu$m and
 is shown in the left panel of Fig.~\ref{fig_temp}. 
The solid line is the {\em face-on} SED of the accretion
disk (i.e.\ the emission in directions within the polar cone, $\theta < \theta_h$) 
and the dashed line is the {\em edge-on} SED  (i.e.\ the emission in directions 
outside the polar cone, $\theta > \theta_h$).  Note that since the SED
along directions that intersect the dusty torus depends on the viewing angle,
for the sake of simplicity we adopt as the {\em edge-on} SED of the
central source the solid-angle average.  Comparing the 
solid and dashed lines in the figure, we note that the SED in the NIR is almost 
isotropical and that a fraction of the optical luminosity from the accretion
disk is able to escape the dusty torus even along the edge-on direction. 
These features are due to the moderate opacity of the torus which limits dust 
self-absorption. Indeed, for the selected model, the optical depth at 
$9.7 \, \mu$m in the equatorial direction ($\theta=90^\circ$) and assuming
a homogeneous dust distribution in the torus is $\tau_{9.7} = 0.1$. This is 
evident also from the prominent silicate emission feature at $10 \mu$m. At longer
wavelengths, the SED of the central source rapidly drops. \\

\noindent
{\it Model B:} \\
\noindent
The second model for the central source was taken from the CAT3D models 
of H\"oenig \& Kishimoto (2010), where Monte Carlo simulations of individual 
clouds are combined with a statistical approach to compute the torus 
emission\footnote{CAT3D  models are available at: {\tt http://www.sungrazer.org/CAT3D.html}.}. 
CAT3D models do not include the direct light from the accretion disk, but 
only the radiation that has been absorbed/emitted or scattered by the dusty
torus. Following Leipski et al. (2013), we added a template to model
the accretion disk emission in the UV/optical band. We adopt the Stalevski et al. (2012) 
spectrum for the {\em face-on} direction, with luminosity scaled to
the UV/optical data. We note however that scaling a CAT3D model to the 
observations of a QSO of known distance already implies an estimate of its 
bolometric luminosity. Also, the accretion disk template of H\"oenig \& 
Kishimoto (2010) is different from that of Stalevski et al. (2012); we will 
comment upon these issues in section~\ref{sec:results}. We selected the dusty 
torus model that, together with the accretion disk template, better fits the data at 
$1\, \mu\mathrm{m} \le \lambda \le 10 \,\mu\mathrm{m}$. The result is shown in the
central panel of Fig.~\ref{fig_temp}. The dusty torus model is characterized by a half-opening
angle of $\theta_h=60^\circ$. However, since in these models the opening 
angle is not defined as a sharp boundary between the 
dust-free polar cone and the torus, but rather as a transition parameter, 
both the {\em face-on} and the {\em edge-on} SEDs are derived by averaging 
over the solid angle within and outside the polar cone. 
The figure shows that in this second model for the central source 
(hereafter, {\em model B}) the emission 
is less isotropic than in the previous case. Compared to model A,
the radiation along edge-on directions at $\lambda < 1 \, \mu$m is negligible.
Though this might be due to the missing information about the attenuated
accretion disk light in CAT3D spectra, it is compatible with the larger opacity of the
selected dusty torus, which is characterized by a mean optical depth of 
$\tau_{9.7} \sim 17$ in the equatorial direction. Accordingly, we assumed 
complete obscuration - i.e.\ no {\em edge-on} emission - in our optical/UV
template.\\

\begin{figure*}
\sidecaption
\includegraphics[width=12cm]{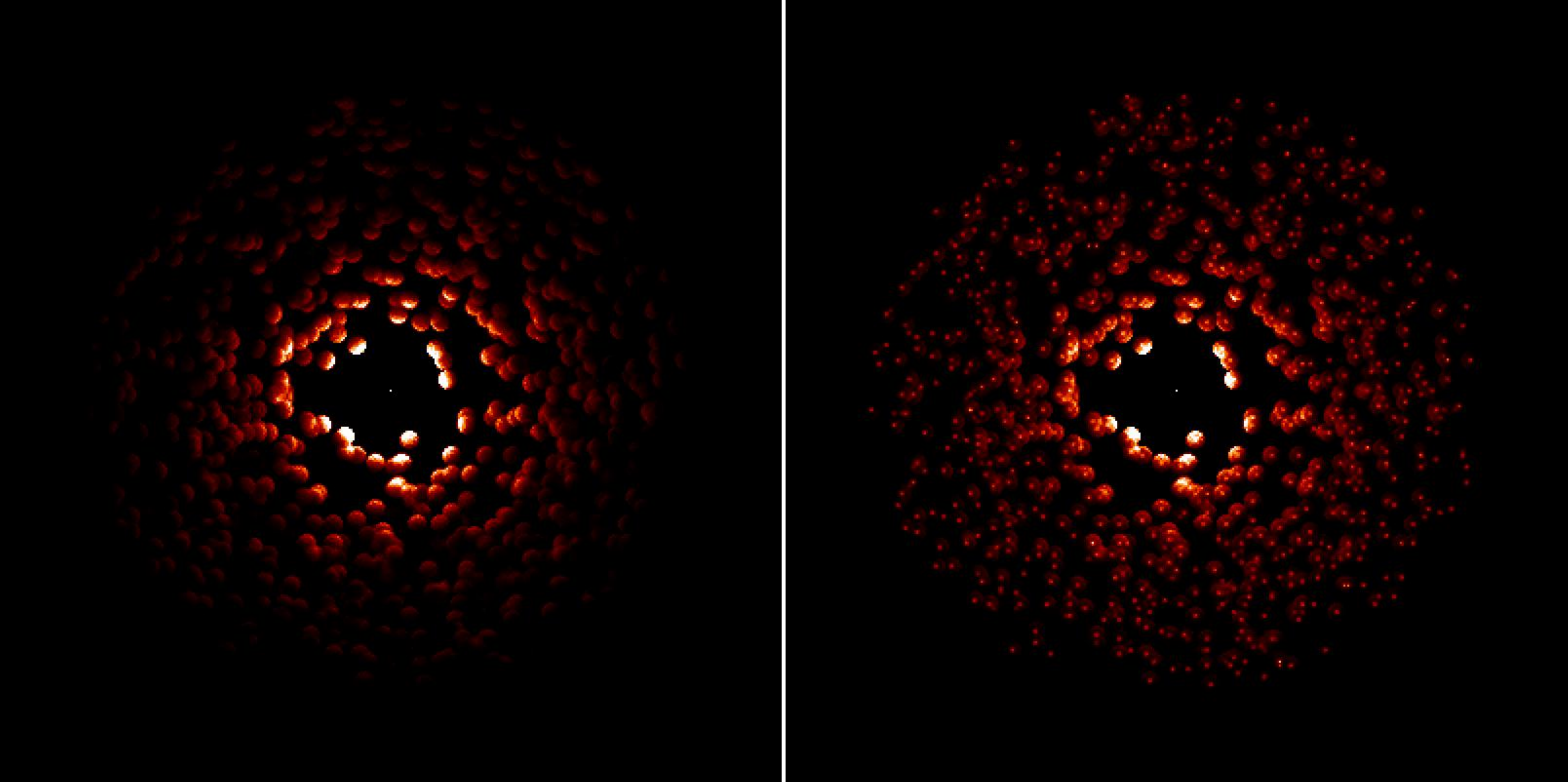}
\caption{Maps of 7 kpc x 7 kpc regions 
representing the radiation emitted at 40$\mu$m in the face-on direction for model A.
In the left panel the only source of radiation is the central one. The right panel 
shows the full model where both the central source and the stars embedded in the 
clouds contribute to dust heating.}
\label{fig_maps}
\end{figure*}

\noindent
{\it Model C:}\\
\noindent
The central source templates presented so far have been selected to 
reproduce the NIR bump at $1\,\mu\mathrm{m} 
\le \lambda \le 10\, \mu\mathrm{m}$ with torus emission. 
However, the exact nature
of this emission poses a challenge to most torus models\footnote{This
problem is inherent to clumpy models (see e.g. Deo et al. 2011), as smooth models 
reproduce this observed feature. 
It also depends significantly on the assumed primary source (Feltre et al. 2012).}
and it is often taken into
account by including an additional hot dust component at the inner edge of the
torus (H\"oenig 2012). In view of this, Leipski et al. (2013, 2014) model the NIR
bumps with hot dust blackbody spectra and select torus models that fit the observed
data at longer wavelengths, in the MIR. Following this strategy, we select a third model
for the central source from H\"oenig \& Kishimoto (2010) templates, that we show
in the right panel of Fig.~\ref{fig_temp} (hereafter, {\em model C}). We do not 
include, however, the component for the NIR bump; we will discuss the effects of this 
omission in section~\ref{sec:results}.
The {\em face-on} and {\em edge-on} SEDs have been obtained as for the second model
discussed above and for the same torus half-opening angle of $\theta_h = 60^\circ$. Note that
the two dusty torus models selected from the H\"oenig \& Kishimoto (2010) templates 
(models B and C)
have the same mean optical depth and cloud properties. The only difference is the adopted
power law index of the radial dust cloud distribution, which is flatter in model C.
This is consistent with having a stronger emission 
at longer wavelengths as the clouds are expected to be colder, on average. 

The derived non-ionizing luminosities of the three models for the central source, that 
will be used for the radiative transfer calculations, are $\rm L_\mathrm{central}=$7.0, 6.2 
and 5.4 $\times 10^{13}$ L$_{\sun}$ for model A, B, and C, respectively.\\

\section{The stellar spectral energy distribution}
\label{sec:stars}

As explained in section \ref{sec:model}, the properties of the host galaxy
are inferred from the output of the semi-analytical code 
\textsc{GAMETE/QSOdust} (Valiante et al. 2011, 2012, 2014). This code
allows to simulate a large number of independent hierarchical histories
for the DM halo that host J1148 at $z = 6.4$. Hence, the values
for the star formation rate and total stellar mass for the
selected model that we have quoted in Section \ref{sec:model} and
that we report here for convenience represent the average among 50
independent merger trees, with the errors showing the associated 1$\sigma$
dispersion. 

At $z = 6.4$ the selected standard IMF model predicts a star formation rate of $\rm SFR = (895 \pm 800 )\, M_\odot/yr$ and a  
total stellar mass of $\rm M_{\star} = ( 3.8 \pm 1.2) \times 10^{11} \, M_\odot$
(Valiante et al. 2011, 2014).
The large dispersion around the mean SFR is due to 
sensible variations of the merger rates and the effects of AGN feedback predicted by the different hierarchical histories.

To compute the SED of the stars in the host galaxy of J1148 we use the public
code P{\'E}GASE v2.0\footnote{http://www2.iap.fr/users/fioc/PEGASE.html} (Fioc \& Rocca-Volmerange 1997)
that allows to reconstruct the SED of stars that form with arbitrary star formation
history, initial mass function and metallicity. The result is shown in the right panel of Fig.\ref{fig_temp},
where we plot the SED produced by the average star formation history (solid black line) and the  shaded region
illustrates the range of SED associated to star formation histories which lie within 1-$\sigma$ of the average.

Even though all the star formation histories predict a peak in the star formation rate at $7 < z < 9$ (see Fig. 4 in Valiante et al. 2014), the $z = 6.4$ 
SED is largely dominated by stars formed in the last $< 100$~Myr of the evolution of the host, because the
 luminosity of the stars formed at the peak of star formation has already faded away.
As expected for an optically bright QSO, Fig.\ref{fig_temp} shows that the radiation emitted by the stars
at optical and NIR wavelengths is always largely subdominant with respect to that produced by the
accretion disk and dusty torus. 

In the radiative transfer model, stellar radiation is assumed to be emitted isotropically. The non ionizing 
luminosity is $\rm L_\mathrm{\star}=(1.5 \pm 1.3) \times 10^{13} L_{\sun}$,  for the mean and $\pm 1 \sigma$
SFRs, respectively.
The adopted spatial distribution for the gas, dust and stars in the
host galaxy will be presented in the next section.

\begin{figure*}
\includegraphics[width=\hsize]{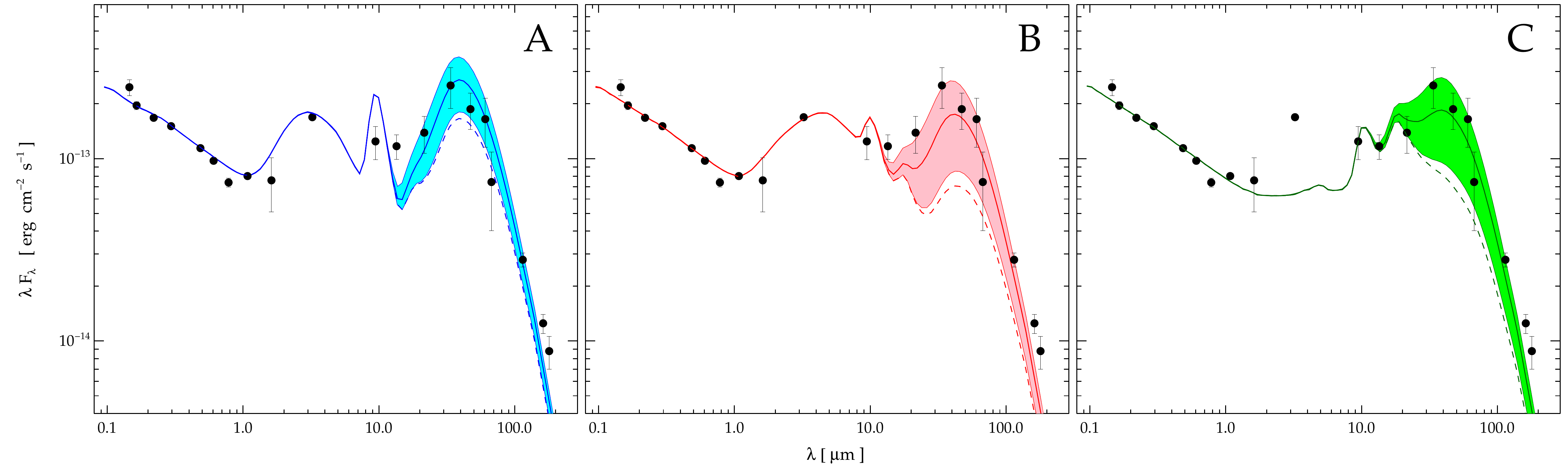}
\caption{The SED of J1148 obtained by the radiative transfer calculation for the different 
inputs for the central source (models A, B and C) presented in Fig.\ref{fig_temp}.
The solid lines show the results including the central source and the stellar component, with the
shaded region illustrating the variations of the stellar SED induced by the 1$\sigma$ scatter of the SFR 
(see the left panel of Fig.~\ref{fig_temp}). The dashed lines show the results obtained using the central source only.
}
\label{fig_seds}
\end{figure*}

\section{The dust distribution in the host galaxy}
\label{sec:dust}

The radiative transfer is computed after distributing dust within a cubic volume centered
on the central source. Two dust-free regions are included in the cube:
{\it (i)} a central sphere of radius 600 pc, since the adopted SED for the central source already takes 
into account radiative transfer through the dusty torus at these scales\footnote{
For a bolometric luminosity of $10^{11}$ L$_{\sun}$, Stalevski et al. (2012) adopt an external 
radius of the dusty torus of 15 pc. Scaling to the bolometric luminosity of J1148, $\rm L_{bol} = 1.6 \times 10^{14}$ L$_{\sun}$
and assuming flux conservation, the radius is $\rm \propto L_{bol}^{1/2}$ and we infer a radius of 600 pc.};
{\it (ii)} a cone that is the
extension of the opening cone of the dusty torus, $ 2 \theta_h$, so that the radiation exiting the dusty
torus unabsorbed is not further absorbed by the host and maintain the observed
spectral slope in the optical. Note that we also explore the possibility
that the polar axis of the dusty torus is inclined by an angle ${\it i} > 0$ with respect to the
polar axis of the host galaxy (see Fig.~\ref{fig:scheme}). Yet, since J1148 is classified as a 
Type-1 QSO, $i << \theta_h$ so that most of the lines of sight to the accretion disk are dust-free.

At $z = 6.4$ the selected model predicts a molecular gas mass
of $\rm M_{H_2} = (3.1 \pm 2.8)\times 10^{10} \, M_\odot$ and a total dust mass of $\rm M_{dust} = (3.2 \pm 2.9)\times 10^8 \, M_\odot$.
Since the dominant contribution to dust enrichment comes from grain growth in dense 
molecular clouds, almost all of the final dust mass
is associated to molecular gas (Valiante et al. 2014). 
Hence, we distribute the dust mass solely in spherical, homogeneous clouds simulating dust 
associated to molecular gas.
As in Bianchi (2008), we adopt a cloud radius of about 100 pc, which is covered by 
about 5 cells of the adaptive grid in the standard resolution settings. With this
clumpy distribution, we minimize the extinction of radiation from the central source
and maximize that from the stellar component since, as we will see below, we 
assume the stars to be embedded in the clouds.

Interferometric observations of J1148 reveal CO emission within 2.5 kpc from the QSO's
center, which corresponds to about $\rm 2.3 \times 10^{10}~M_{\odot}$ of molecular hydrogen 
(Walter et al. 2004). From these observations, we derive a molecular gas surface density
of about $\rm 1200~M_{\odot}/pc^2$, which we assume to represent the surface density
of the individual clouds. Such high surface density clouds are not unrealistic for QSO host galaxies
at high redshifts. In fact, CO excitation analyses show that the observed transitions can
be modeled with a single gas component with density $\rm n_{H_2} \sim 10^{3.6} - 10^{4.3} cm^{-3}$
and temperature $\rm T \sim 40 - 60~K$, suggesting that the emission comes from a very compact and
central region of the host (see Carilli \& Walter 2013 and references therein). 

Clouds are distributed homogeneously within a radius of 3.0 kpc, so that the average
surface density within 2.5 kpc from the QSO's is similar to what observed.
To be consistent with the radiative transfer models for the central source, we use the 
Milky Way (MW) dust composition of Draine (2003). For simplicity, we consider the mean extinction and emission properties
of the models, i.e.\, we adopt a single grain of mean properties, rather than a full distribution
of grain sizes and materials. Also, we assume thermal equilibrium emission. We will discuss
the effects of these assumptions later.

Given the average molecular gas mass predicted by the model, we end up by having $\sim 850$ clouds, 
each one having a molecular gas mass of $\rm 3.6  \times10^{7}~M_{\odot}$. 
The average gas to dust ratios of the clouds is $\rm G/D \sim 100$, 
comparable to what measured for the moderately dense interstellar medium of the Milky Way (Jenkins 2009).
As a result, the optical depth of the clouds  along the diameter in the V-band is very large,  $\rm \tau_V \sim 99$
for Draine (2003) dust composition. Thus, some edge-on lines of sight 
could be severely attenuated in the optical-MIR. 


Finally, stellar radiation is assumed to be emitted at the center of each cloud. Given the
high optical depth of clouds, the stellar radiation is entirely 
absorbed within each cloud. Thus, the stellar contribution to dust emission in all the models
should be considered as an upper limit.

\section{Results and discussion}
\label{sec:results}

A qualitative picture of the results of the RT calculation is presented in Fig.~\ref{fig_maps}, which shows maps
of the radiation emitted at 40$\mu$m in the face-on direction for model A. Due to the
large optical depth of the dusty clouds in the host galaxy, the central source is able to illuminate and heat  
only the surface of the closest clouds  (left panel); the most distant clouds 
appear to contribute to the FIR only when the dust is heated by the stellar radiation
(right panel).

More quantitatively, the full SED of J1148 in the face-on direction is shown in Fig.~\ref{fig_seds},
where each panel represents the results of the RT calculation adopting one of the 
central source models described in Section \ref{sec:central}. The solid lines 
show the predicted SED when both the radiation from the central source and 
from the stars contribute to heat the dust in the host galaxy, powering the FIR emission. As
indicated by the shaded regions, which account for the 1$\sigma$ variation on the SFR
among different hierarchical histories of the host galaxy 
(see the right panel of Fig.~\ref{fig_temp}), the radiation from stellar sources affects the
SED only at $\lambda > 10~\mu$m; at lower wavelengths, the emission is entirely dominated by the central source.
Even though it is not possible to separate the two contributions from the FIR SED (the dust 
emission depending in a non-linear way on the radiation absorbed by the dust from each type of
sources), we can have a clue at their relative importance by comparing the above results with the RT
calculation  when the dust in the host galaxy is heated only by the central source (dashed line in each panel).

For model A, the observed FIR emission is marginally consistent with dust heated by the central source only. 
Integrating\footnote{As customary for
observations, here we neglect that the SED might be anisotropic and obtain the luminosity 
integrating over the whole solid angle (i.e.\ we multiply the integral under the SED by $4\pi D_L^2$).} the dashed line 
shown in the left panel of Fig.~\ref{fig_seds}, we find that the luminosity in the 
wavelength range $\rm [20 -1000] \, \mu$m is $\rm L_{FIR,central} = 
2.3 \times 10^{13} L_\odot$, where instead the luminosity of the full model 
including stellar sources (shaded region) is in the range
$\rm L_{FIR,TOT} = (2.5 - 4.7)\times 10^{13} L_\odot$.
Comparing these figures with the FIR luminosity inferred from the observational data, 
$\rm L_{FIR, obs} = 3.3 \times 10^{13} L_\odot$ (see section \ref{sec:model}), we conclude
that $\rm L_{FIR,AGN} =  0.7 \, L_{FIR, obs}$ and that the radiation emitted by the central source
in model A dominates the heating of the dust which powers the FIR emission.

For model B (central panel of Fig.~\ref{fig_seds}) the contribution of the 
central source to the FIR SED is smaller, but still significant: we find
$\rm L_{FIR,central} = 1.1 \times 10^{13} L_\odot$ (dashed line)
and $\rm L_{FIR,TOT} = (1.3 - 3.7) \times 10^{13} L_\odot$ (shaded region).
Thus, we conclude that the central source contribute to about $30 \%$ of the
FIR emission, $\rm L_{FIR,AGN} =  0.33 \, L_{FIR, obs}$.

Results for model C (right panel) appear to be similar, with 
$\rm L_{FIR,central} = 1.5 \times 10^{13} L_\odot$ (dashed line)
and $\rm L_{FIR,TOT} = (1.7 - 4.1) \times 10^{13} L_\odot$ (shaded region). 
Hence, we find that $\rm L_{FIR,AGN} =  0.45 \, L_{FIR, obs}$. However,
in this case, the contribution of the central source to the observed 
FIR emission is not only the result of dust heating in the host galaxy, but also directly
through its intrinsic, unattenuated, SED. For our particular choice among
the H{\"o}nig \& Kishimoto (2010) models and normalization, the direct 
light accounts for more than half of the central source contribution 
to the FIR output. However, because of the rapid decrease of the SED with wavelength,
alternative choices for model C (and different normalizations) might result
in widely different contributions of the central source to the FIR SED.
For example, in the model chosen by Leipski et al. (2013), the FIR output
of direct light from the central source is about 20\% of the total, similar
to our choice here; in the updated model of Leipski et al. (2014), the contribution is only 5\%.

The results of the RT calculation show that the contribution of
dust heating from the central source can range from moderate to significant, $\rm L_{FIR,AGN} =  (0.33 - 0.70) \, L_{FIR, obs}$
depending on the adopted models and templates for the central source.
This affects the determination of the SFR from the FIR luminosity and
our ability to constrain the connection between the BH and its host
galaxy from the IR properties of QSOs.
In what follows, we discuss some critical aspects of the radiative transfer model and 
how these might eventually affect the results presented above. \\

\begin{figure*}
\center
\includegraphics[width=\hsize]{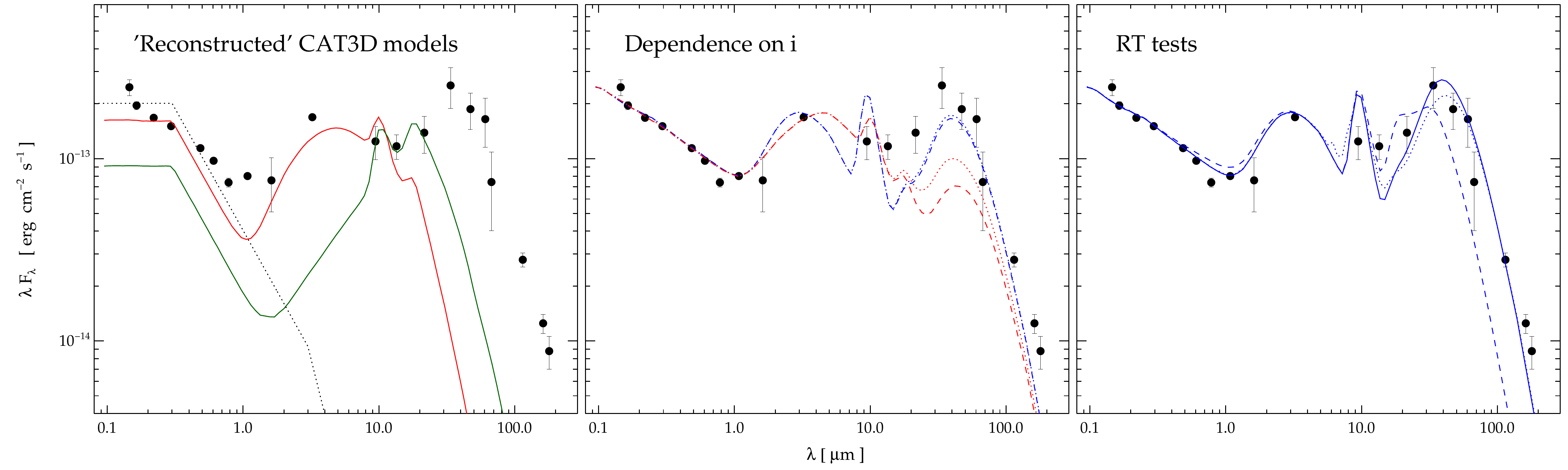}
\caption{{\it Left panel}: the reconstructed full {\em face-on} SED of the adopted CAT3D models,
adding to the dusty torus SED (red solid line for model B and green solid line for model C) 
a dust-free view of the  H\"onig \& Kishimoto (2010)
intrinsic SED of the accretion disk. 
The dotted line is the SED of the accretion
disk scaled to the observed optical/UV datapoints. 
{\it Central panel}:
SEDs for models A (blue) and B (red)  using the central source only and adopting two 
 different 
inclination angles of the polar axis of the torus with respect to that of the host: $i = 0$ (dashed lines) and $20^\circ$ (dotted lines).
{\it Right panel}: 
SEDs for model A but with dust mass reduced by a factor of 10
(dashed line) and by using the original dust mass but
increasing the resolution and adopting a full grain size distribution plus 
stochastic heating (dotted line). The solid line is the reference model 
from the left panel of Fig.~\ref{fig_seds}.}
\label{fig_test}
\end{figure*}

\noindent
{\it Limitations and uncertainties of the central source models}\\
\noindent
Among the many different models proposed in the literature, 
here we have explored three different templates for the radiation 
emitted by the dusty torus that surrounds the 
nuclear BH accretion disk. From a methodological point of view, the Stalevski et al. (2012) 
library, out of which we have selected model A, is better suited to be
used as a central source of radiation in our RT model. In fact, it
includes the radiation emitted by the accretion disk,
so that the predicted dust-free face-on SED can be easily normalized 
to match the observed UV/optical data points (see section \ref{sec:central}). 
Conversely, using the CAT3D library, out of which we have selected 
models B and C, is more challenging. This can be seen in the left panel of Fig.~\ref{fig_test}, where
we have reconstructed the full {\em face-on} view of the original CAT3D models, 
adding to the selected dusty torus SED (models
B and C) the original template used by H\"oenig \& Kishimoto 
(2010) for the accretion disk. It is evident that the original accretion disk template
does not show the same spectral slope as the data. H\"oenig \& Kishimoto (2010) 
claim that in their models the actual shape of the spectrum is much less important than the 
total luminosity. Indeed, scaling their accretion disk template to the optical/UV
data in the range $\rm [0.1 - 1]~\mu$m (dotted line),  we derive a bolometric 
luminosity of $1.0 \times 10^{14}$ L$_{\sun}$, which falls within the estimates reported 
in section~\ref{sec:sed}. However, {\it this is not the intrinsic bolometric luminosity
corresponding to the dusty torus template models B and C shown in Fig.~\ref{fig_seds}}. In fact, 
scaling the CAT3D models to observations of a QSO of known luminosity distance
allows to determine the physical scale of the sublimation radius (i.e.
the inner edge of the dusty torus) and thus the bolometric luminosity
intrinsic to each template\footnote{See Sect. 4.1 of H\"oenig \& Kishimoto (2010) 
and the documentation on the dedicated webpage.}. For the dusty torus model B,
the bolometric luminosity estimated in this way is smaller, $8.2 \times 10^{13}$ L$_{\sun}$. 
The reconstructed face-on SED corresponding to
this luminosity (solid line in the left panel of Fig.~\ref{fig_test}) is marginally consistent with the optical/UV data. 
Note that now the dusty torus model B does not reproduce the NIR data as it used to in the central panels of Fig.~\ref{fig_temp} and \ref{fig_seds}. 
This is because model B  was originally selected in combination with the UV/optical template of 
Stalevski et al. (2012), which - following the flatter slope of the observed
optical/UV data - extends to 5 $\mu$m and contributes to the NIR.

More critical is the problem for the dusty torus model C (dashed line in the left panel of Fig.~\ref{fig_test}). 
The bolometric luminosity derived, after scaling to match 
the MIR data, is $5 \times 10^{13}$ L$_{\sun}$, definitely lower than what
predicted by observations. Conversely, if the model was scaled to the
optical/UV observations (i.e. if the dashed line in Fig.~\ref{fig_test}
was made to coincide with the dotted line), the MIR prediction would be
higher than the data\footnote{Browsing the
Stalevski et al. (2012) dataset we find similar results:
if we assign the models the same bolometric luminosity as derived from
the optical/UV data, then models peaking in the NIR might reproduce 
the observed NIR data, while models peaking in the MIR have a larger
luminosity.}.
   
The results shown in the left panel of Fig.~\ref{fig_test} suggest that choosing independent templates for the
optical/UV emission of the accretion disk and for the dusty torus, as we have done
for the central source models B and C, does not ensure that the heating source
in the RT model is compatible with the observations. Note that this same 
approach has been recently followed by Leipski et al. (2013, 2014) and was
applied by Barnett et al. (2015) to the SED of the most distant QSO at $z = 7.1$.

We also note that  a similar problem
affects the blackbdody template used by Leipski et al. (2013) to fit the observed
NIR peak of high-$z$ QSO spectra (see also footnote 4 in Section 4), which is thought to arise from the radiation 
emitted by hot dust at the boundary between the accretion disk and the torus: 
in a consistent RT model, this emission should be included as an additional heating source, 
and would thus alter the dusty torus SED in ways that are difficult to predict 
without running a dedicate RT computation. 

Based on the above considerations, {\it we can reject model C}
and conclude that only dusty torus SEDs peaking in the NIR (and 
thus reproducing the observed bump) are acceptable, in that 
they are based on intrinsic accretion disk luminosities which can
be reconciled with the observations. \\

\noindent
{\it Dependence on the inclination angle i}\\
\noindent
The results discussed above do not change significantly if we allow for a moderate disalignment between the polar axis of the 
dusty torus and the polar axis of the host galaxy. This is illustrated by the central panel of Fig.~\ref{fig_test},
where we show the predicted SEDs for models A and B using the central source only and adopting two 
different inclination angles. 
When $i = 20^\circ$ (see also the geometric scheme in Fig.~\ref{fig:scheme}),
the FIR luminosity contributed by the central source in model A increases only slightly, 
to $\rm L_{FIR,central} = 2.4 \times 10^{13} L_\odot$. This is a consequence of the relatively low optical depth 
of the dusty torus in this model. Indeed, the left panel of Fig.~\ref{fig_temp} shows that there is a significant fraction 
of the radiation emitted by the
accretion disk that is able to escape the dusty torus in the edge-on direction both at MIR and optical wavelengths; 
these two components almost equally contribute to dust heating and their effect is not sensitive to the inclination angle
of the dusty-torus. In fact, when ${\it i}  > 0$ the dusty clouds in the host galaxy will be exposed to a smaller fraction of the 
radiation escaping the torus in the edge-on direction, $\rm1 - sin^2({\it i}/2)/(2-sin^2\theta_h)$, but - at the same time - 
will intercept a fraction $\rm sin^2({\it i}/2)/(2-sin^2 \theta_h)$ of the radiation emitted by
the accretion disk in the face-on direction, that is unattenuated by the dusty torus. These two components partly compensate,
leading to a negligible effect of the inclination angle on the FIR emission. Given the larger optical depth of the dusty torus, in model B 
the net effect of the disalignment $i$ is larger (see the central panel of Fig.~\ref{fig_test}).
The unattenuated UV/optical radiation from the accretion
disk now intercepts the dusty clouds distributed within a solid angle  $\rm sin^2({\it i}/2)/(2-sin^2 \theta_h)$. As a result,
the contributions of the central source to the FIR luminosity increase to $\rm L_{FIR,central} = 
1.5 \times 10^{13} L_\odot$, contributing to $0.45$  of
 $\rm L_{FIR, obs}$.\\

\noindent
{\it Dependence on the dust mass and composition}\\
\noindent
In the right panel of Fig.\ref{fig_test}, we show the predicted SED assuming the same heating
sources as for model A, but with a factor 10 smaller 
dust mass, $\rm 3.1 \times 10^7 M_{\odot}$ (dashed line).
In this case, both the amplitude and frequency of the
thermal dust emission associated to the host galaxy shift to lower values as the clouds are more
transparent and the dust temperature is larger.
Hence, although dust mass estimates derived by fitting
the observed FIR emission are plagued by a number of uncertainties, our RT 
model shows that published values have not been severely over-estimated.
Similarly, the mass-averaged dust temperatures computed by the RT calculations
are $\rm <T_d> =  56 \, and \, 52~K$ for models A and B, respectively; these 
values are very close to the dust temperature estimated from the observed data
points assuming optically thin dust emission (see section \ref{sec:model}). 

Throughout the analysis, we have adopted the average Draine (2003)
dust model for which, $k(\lambda) = k_0 \,(250\, {\rm \mu m}/\lambda)^\beta$ with $k_0 = 4 \, \rm cm^2/gr$ and 
$\beta = 2.08$ (Bianchi 2013). Even at FIR wavelengths, some of the 
features predicted by the RT calculation may be affected by the adopted dust properties.
Indeed, none of the three models shown in Fig.~\ref{fig_seds} can fully account for the
observations at $\lambda \ge 100 \, \mu$m. The excess emission could be accounted for
by dust emissivity with a flatter spectral index, as predicted by SN dust models (Bianchi \& Schneider 2007) 
and as suggested by the fits of the FIR modified blackbody 
in Leipski et al. (2013). However, the same excess could be the result of dust further 
out of the source, or passively heated clouds (without sources in them),
which we have not considered. Note also that, by analyzing the optical-near infrared spectra of 33 quasars with redshifts $ 3.9 \leq z \leq 6.4$,
Gallerani et al. (2010) find that their mean extinction properties deviate from that of the SMC, which 
has been shown to reproduce the dust reddening of most quasars at $z < 2.2$. This may be 
an indication of different dust properties
at high redshift.

Finally, we have checked that increasing the resolution by a factor 4 (100 pc covered by
about 20 cells of the adaptive grid) does not affect our results.
Similar conclusions hold for models where we have taken into account the full distribution of grain sizes
and materials (rather than a single grain with mean properties), and the effects of
stochastic heating (rather than thermal equilibrium emission). All the test runs
show differences with respect to our reference models
that are smaller than the photometric errors at $\lambda > 30 \mu$m (see the dotted line in the right panel
of Fig.\ref{fig_test}).

\section{Conclusions}
\label{sec:conclusions}

One of the main motivations of the present study was to explore to what extent the far infrared
properties of high-$z$ QSOs can help constrain current scenarios for the coevolution of the
BH and its host galaxy in the first Gyr of cosmic evolution. 

In FIR luminous high-$z$ QSOs, such as the $z = 6.4$ QSO SDSS J1148 that we have selected as
a representative case, the observed FIR emission is generally attributed to an ongoing strong
starburst in the host galaxy, with inferred star formation rates that can range from $\sim 1000 \,\rm M_\odot/yr$ up to 
$\sim 3700\, \rm M_\odot/yr$ (see Table 1 in Valiante et al. 2014). Yet, these figures can only provide a loose upper
limit on the star formation rate. 
In fact, by means of a detailed RT calculation, we show that:

 \begin{itemize}
 \item the radiation emitted by the central source,
which dominates the observed SED from the UV/optical to near and mid-infrared wavelengths,  
provides an important source of heating for the dust distributed in the host galaxy, powering
at least 30 and up to 70\% of the observed far infrared emission. 

\item the FIR SED of SDSS J1148 can only be reproduced by models where the star formation
rate in the host galaxy at $z = 6.4$ is as large as $\rm SFR \sim 900 \, M_\odot/yr$, as predicted by 
the selected standard IMF model (Valiante et al. 2011).  In fact, a top-heavy IMF model, with reduced star formation efficiencies,
predicts a SFR at $z = 6.4$ of only $\rm 66 \pm 59 \, M_\odot/yr$, too small to supply the required heating,
in addition to that of the central source, and to power the observed FIR emission. 

\end{itemize}

Hence, we conclude that our analysis supports a scenario for the co-evolution of the first SMBHs
and their host galaxies where, during their hierarchical assembly, the evolutionary
tracks followed by the systems in the $\rm M_{BH} - M_{star}$ plane approach the local relation
from the bottom (see Fig.~5 in Valiante et al. 2014): during the first period of evolution, at redshifts
$8-10 < z < 30$, in all the progenitor galaxies stars form according to a standard,
Salpeter-like IMF via quiescent star formation and merger-driven bursts. At the same time, the 
nuclear BHs grow through mergers with other BHs and gas accretion; hence
 BHs and their host galaxies grow hand-in-hand, in a symbotic way (Volonteri 2012).  At
 $5 - 6.4 \le z \le 8 - 10$, however, the BH starts growing faster than the stellar bulge and the predicted 
 $\rm M_{BH} - M_{star}$ evolution progressively steepens, falling within the observed scatter of
 the local relation. This scenario implies that current dynamical
mass measurements may have missed an important fraction of the
host galaxy stellar mass.
 
It is important to note that the transition between the starburst-dominated stage and the QSO-dominated stage of the evolution
is regulated by the effects of BH feedback: for the QSOs that we have investigated, we find that when the nuclear BH masses 
have grown to $\rm \sim 2 \times 10^8 - 10^9 M_\odot$, strong AGN-driven outflow rates of 
$\rm (4 - 6) \times 10^3 M_\odot/yr$ damp the star formation rate and un-obscure the line of sight 
towards the QSO. At these large outflow rates,
the host galaxies would be completely depleted of their gas content in less than 20 Myr, shutting down 
both the star formation rate and the BH activity; as a result, we predict the active and bright QSO phase to 
last $\rm \sim 10^7 yr$ (Valiante et al. 2014). During this relative short phase, the FIR luminosity is expected to 
respond to the rapidly changing physical conditions in the host galaxy. 
Our analysis suggest that FIR-luminous QSOs are likely tracing 
a phase of the evolution where the gas content in the host galaxy can still sustain strong starbursts, 
with $\rm 100 M_\odot/yr < SFR < 1000 M_\odot/yr$. When the conditions in the ISM are such that the 
star formation rate fall below $\rm < 100 M_\odot/yr$, the FIR emission can only be powered by dust 
heating from the central source, and we expect a stronger correlation between the UV/optical, NIR/MIR
and the residual FIR luminosities. 

It is interesting to note that this results resonate with the recent analysis by Leipski et al. (2013, 2014), 
where the authors find that the average SED of QSOs $z > 5$ with {\it Herschel} detections, including SDSS J1148, 
are preferentially found at the high luminosity end of their sample (for $\rm L_{UV/opt}$ and in particular
for $\rm L_{NIR}$). Moreover, as discussed in section \ref{sec:intro}, they also find that some optical and
MIR luminous QSOs do show FIR emission and others do not. This is an indication that star formation
is the dominant driver for the additional FIR component observed in their {\it Herschel}-detected sources.
Following a similar approach, Barnett et al. (2015) has recently analyzed the SED of 
QSO ULAS J1120+0641 at  $z = 7.1$. This object has been detected by {\it Herschel} PACS at
$100$ and $160 \, \mu$m but not by {\it Herschel} SPIRE at $250$,
$350$ and $500 \, \mu$m, hence it would be a member of the 
partly ({\it Herschel}) detected objects, according to the classification by Leipski et al. (2014). On average,
these sources are optically luminous AGN with strong NIR and MIR emission, but do not show a bright FIR emission,
consistent with the idea that the central source is likely contributing significantly, if not dominantly, to the FIR emission.
Using local scaling relations, Barnett et al. (2015) estimate a star formation rate in the range $\rm 60 - 270 \, M_\odot/yr$
from the [CII] line luminosity and the $\rm 158\, \mu$m continuum luminosity and conclude that, at the time
of observation, the BH was growing in mass 100 times faster than the stellar bulge. This is consistent with the idea that
the source was likely caught in action when the QSO-driven wind has already significantly depleted the host galaxy of its
gas content, suppressing star formation. 

Our results are in good agreement with another study done on the same object by 
Li et al. (2008). By applying a 3D radiative transfer code to a hydrodynamical simulation, these authors 
show that the observed SED and inferred dust properties of SDSS J1148 request vigorous star formation in the 
merging progenitors and a substantial contribution to dust heating and the FIR luminosity from the AGN. 
They also suggest that the quasar host should have already built up a large stellar population by $z \sim 6$, 
in agreement with our best-fit evolutionary scenario described above. 
Our study, which is based on a semi-analytical code, complements these previous findings allowing an 
exploration of a larger parameter space (in terms of stellar IMF and star formation history) and a 
more sophisticated treatment of the chemical enrichment of the ISM by dust and metals.

\begin{acknowledgements}
We thank the anonymous Referee for her/his insightful comments, Sebastian H\"onig, Christian Leipski and Dominik Riechers for the kind clarifications about their work.
RS and RV acknowledge the support and hospitality of the INAF/Osservatorio Astrofisico di Arcetri during part of the work.
The research leading to these results has received funding from the European Research Council under the European Union's
Seventh Framework Programme (FP/2007-2013) / ERC Grant Agreement n. 306476.
S.Salvadori acknowledges the support from the 
Netherlands Organisation for Scientific Research 
(NWO), VENI grant 639.041.233
\end{acknowledgements}


\begin{thebibliography}{99}

\bibitem{}
Baes M., Verstappen J., De Looze I., Fritz J., Saftly W., Vidal P\'erez E., Stalevski M., Valcke S., 2011, ApJS, 196, 22

\bibitem{}
Barnett R., Warren S. J., Banerji M., McMahon R. G., Hewett P. C., Mortlock D. J. et al. 2015, A\&A, in press 

\bibitem{} 
Barth A.J., Martini P., Nelson C.H., Ho L.C., 2003, ApJ, 594, L95



\bibitem{}
Bianchi S.,  2008, A\&A, 490, 461


\bibitem{}
Bianchi S., Schneider R.,  2007, MNRAS, 378, 973

\bibitem{}
Bianchi S.,  \& Xilouris E.~M.,  2011, A\&A, 531, L11

\bibitem{}
Bianchi S., 2013, A\&A, 552, 89

\bibitem{}
Carilli C. \& Walter F. 2013, ARAA, 51, 105

\bibitem{}
Cicone C., Maiolino R., Gallerani S., Neri R., Ferrara A., Sturm E. et al. 2015,  A\&A , 574, 14

\bibitem{}
Deo R. P., Richards G. T., Nikutta R., Elitzur M., Gallagher S. C., Ivezic Z., Hines D. 2011, ApJ, 729, 108

\bibitem{}
Dai Y. S., Bergeron J., Elvis M., Omont A., Huang J., Bock J. et al. 2012, ApJ, 753, 33 

\bibitem{}
de Bennassuti M., Schneider R., Valiante R., Salvadori S. 2014, MNRAS, 445, 3039


\bibitem{}
Draine B.~T.,  2003, ARA\&A, 41, 241

\bibitem{} 
Dwek E., Galliano F., Jones A.P., 2007, ApJ, 662, 927

\bibitem{}
Elbaz D., Hwang H. S. , Magnelli B., Daddi E., Aussel H., Altieri B. et al. 2010, A\&A, 518, L29

\bibitem{}
Fan X. et al. 2003, ApJ, 125, 1649

\bibitem{}
Fan X. et al., 2004, AJ, 128, 515

\bibitem{}
Fanidakis N., Macci\'o A. V., Baugh C. M., Lacey C. G., Frenk C. S. 2013, MNRAS, 436, 315

\bibitem{}
 Feltre A., Hatziminaoglou E., Fritz J., Franceschini A. 2012, MNRAS, 426, 120

\bibitem{}
Feltre A., Hatziminaoglou E., Hern\'an-Caballero A., Fritz J., Franceschini A., Bock J. et al. 2013, MNRAS, 434, 2426


\bibitem{}
Fioc M., Rocca-Volemrange B. 1997, A\&A, 326, 950

\bibitem{} 
Gallerani S., Maiolino R., Juarez Y., Nagao T., Marconi A., Bianchi S., et al. 2010, A\&A, 523, A85

\bibitem{}
Haas M., M\"uller S. A. H., Chini R., Meisenheimer K., Klaas U., Lemke D. et al. 2000, A\&A, 354, 453

\bibitem{}
Haas M., Klaas U., M\"uller S. A. H., Bertoldi F., Camenzind M., Chini R. et al. 2003 A\&A, 402, 87

\bibitem{}
 Hatziminaoglou E., Omont A., Stevens J. A., Amblard A., Arumugam V., Auld R. et al. 2010, A\&A, 518, L33

\bibitem{}
H\"onig S., Proceedings of the Torus Workshop, 2012, held 5 December, 2012 at University of Texas, San Antonio, 2012, p. 157 

\bibitem{}
H\"onig S., Kishimoto M., 2010, A\&A, 523, A27

\bibitem{}
Holwerda B.~W.,  Bianchi S.,  B{\"o}ker T.,  Radburn-Smith D.,  de
  Jong R.~S.,  Baes M. et al.
2012, A\&A, 541, L5

\bibitem{}
Hines, D. C., Krause, O., Rieke, G. H., et al. 2006, ApJ, 641, L85
  
\bibitem{}
Iwamuro, F., Kimura, M., Eto, S., et al. 2004, ApJ, 614, 69
  
\bibitem{}
Jenkins E. B., 2009, ApJ, 700, 1299  

\bibitem{}
Jiang, L., Fan, X., Hines, D. C., et al. 2006, AJ, 132, 2127

\bibitem{}
{Leipski} C.,  {Meisenheimer} K.,  {Klaas} U.,  {Walter} F.,  {Nielbock} M.,
  {Krause} O. et al.
2010, A\&A, 518, L34

\bibitem{}
{Leipski} C.,  {Meisenheimer} K.,  {Walter} F.,  {Besel} M.-A.,  {Dannerbauer}
  H.,  {Fan} X.
  et al.
  2013,
  ApJ, 772, 103L

\bibitem{}
Leipski C., Meisenheimer K., Walter F., Klaas U., Dannerbauer H., De Rosa G. et al. 2014, ApJ, 785, 154

\bibitem{}
Li Y., Hopkins P. F., Hernquist L., Finkbeiner D. P., Cox T. J., Springel V. et al. 2008, ApJ 678, 41

\bibitem{}
Lutz D., Sturm E., Tacconi L. J., Valiante E., Schweitzer M., Netzer H. et al. 2008, ApJ, 684, 853 

\bibitem{} 
Maiolino R., Gallerani S., Neri R., Cicone C., Ferrara A., Genzel R.,
2012, MNRAS, 425, L66



  
 \bibitem{}
 Netzer H., Lutz D., Schweitzer M., Contursi A., Sturm E., Tacconi L. J. et al. 2007, ApJ, 666, 806 


\bibitem{}
Rowan-Robinson M., 1995, MNRAS, 272, 737


\bibitem{}
{Richards} G.~T.,  {Lacy} M.,  {Storrie-Lombardi} L.~J.,  {Hall} P.~B.,
  {Gallagher} S.~C.,  {Hines} D.~C. et al. 
  2006, ApJS, 166, 470


\bibitem{}
Schweitzer M., Lutz D., Sturm E., Contursi A., Tacconi L. J., Lehnert M. D. et al. 2006, ApJ, 649, 79

\bibitem{}
{Stalevski} M.,  {Fritz} J.,  {Baes} M.,  {Nakos} T.,    {Popovi{\'c}} L.~{\v
  C}.,  2012, MNRAS, 420, 2756

\bibitem{}
{Valiante} R.,  {Schneider} R., {Bianchi} S., {Andersen} A. C.  2009, MNRAS,
  397, 1661

\bibitem{}
{Valiante} R.,  {Schneider} R.,  {Salvadori} S.,    {Bianchi} S.,  2011, MNRAS,
  416, 1916


\bibitem{}
{Valiante} R.,  {Schneider} R.,  {Maiolino} R., {Salvadori} S.,    {Bianchi} S.,  2012, MNRAS,
427, L60

\bibitem{}
Valiante R., Schneider R., Salvadori S., Gallerani S. 2014, MNRAS, 444, 2442 

\bibitem{}
Volonteri M. \& Rees M. J. 2006, ApJ, 650, 669


\bibitem{}
{Volonteri} M.,  2012, Science, 337, 544





\bibitem{}
{Walter} F.,  {Carilli} C.,  {Bertoldi} F.,  {Menten} K.,  {Cox} P.,  {Lo}
  K.~Y.,  
  et al. 2004, ApJL, 615, L17

\bibitem{} 
Willott C.J., McLure R.J., Jarvis M.J., 2003, ApJ, 587, L15

\end{thebibliography}
\end{document}